\documentclass[twocolumn]{emulateapj}

\usepackage{amssymb}
\usepackage{epsfig}
\usepackage{epstopdf}
\usepackage{amsmath}
\usepackage{mathrsfs}
\usepackage{amsfonts}
\usepackage{textcomp}
\usepackage{color}

\def\apj{\mbox{ApJ}}
\def\apjl{\mbox{ApJL}}
\def\apjs{\mbox{ApJS}}

\def\mnras{\mbox{MNRAS}}
\def\aj{\mbox{AJ}}

\def\aap{\mbox{A\&A}}

\def\mu{\textmu m }
\def\mus{\textmu m}

\shorttitle{Silicate features in active galaxies}
\shortauthors{Hatziminaoglou et al.}

\begin{document}

\title{A complete census of silicate features in the mid-infrared spectra of active galaxies}

\author{E. Hatziminaoglou$^{1}$}
\author{A. Hern\'an-Caballero$^{2}$} 
\author{A. Feltre$^{3}$}
\author{N. Pi\~nol-Ferrer$^{4}$}

\affil{$^{1}$ESO, Karl-Schwarzschild-Str. 2, 85748 Garching bei M\"unchen, Germany, ehatzimi@eso.org}
\affil{$^{2}$Instituto de F\'isica de Cantabria, CSIC-UC, Avenida de los Castros s/n, 39005, Santander, Spain}
\affil{$^{3}$Institut d'Astrophysique de Paris, 98 bis boulevard Arago, 75014 Paris, France}
\affil{$^{4}$Department of Astronomy, Stockholm University, AlbaNova Center, 106 91 Stockholm, Sweden}

\begin{abstract}
We present a comprehensive study of the silicate features at 9.7 and 18 \mu of a sample of almost 800 active
galactic nuclei (AGN) with available spectra from the {\it Spitzer} InfraRed Spectrograph (IRS). 
We measure the strength of the silicate feature at 9.7 \mus, S$_{\rm 9.7}$, before and after subtracting
the host galaxy emission from the IRS spectra. 
The numbers of type 1 and 2 AGN with the feature in emission increase by 20 and 50\%, respectively, 
once the host galaxy is removed, while 35\% of objects with the feature originally in absorption
exhibit it in even deeper absorption. 
The peak of S$_{\rm 9.7}$, ${\rm \lambda}_{\rm peak}$, has a bimodal distribution when the feature is in 
emission, with about 65\% of the cases showing ${\rm \lambda}_{\rm peak} > 10.2$ \mus.
Silicates can appear in emission in objects with mid-infrared (MIR) luminosity spanning over six orders of magnitude.
The derived distributions of the strength of the silicate features at 9.7 and 18 \mu provide a solid test bed
for modeling the dust distribution in AGN.
Clumpiness is needed in order to produce absorption features in unobscured AGN and 
can also cause the silicates to be in absorption at 9.7 \mu and in emission at 
18 \mu in type 1 sources. 
We find the `cosmic' silicates of Ossenkopf et al. to be more consistent with the observations
than Draine's `astronomical' silicates.
Finally, we discuss the possibility of a foreground absorber to explain the deep silicate absorption
features in the MIR spectra of some type 2 AGN.
\end{abstract}


\section{Introduction}\label{intro}

The InfraRed Spectrograph (IRS; \citealt{houck04}) onboard the {\it Spitzer} Space telescope 
provided, over a period of more than five years, low- and high-resolution mid-infrared (MIR) 
spectra from many thousands of galactic and extragalactic sources,
at wavelengths between 5 and 40 \mus.

In active galactic nuclei (AGN), the MIR emission is believed to be UV light reprocessed by the 
hot dust surrounding the AGN. The dust, often assumed to form a toroidal structure in parsec scales 
around the nucleus in a plain extending that of the accretion disk, is considered to be distributed 
either smoothly \cite[e.g.][]{pier92, granato94, fritz06} or in clumps \cite[e.g.][]{hoenig06,nenkova08}.
Dust is believed to consist of graphite and silicate grains, each leaving their unmistakable signature in
the Spectral Energy Distribution (SED) of AGN, namely the $\sim$1500K black-body 
like rise of the MIR continuum of type  (unobscured) AGN \cite[e.g.][]{hatzimi05}, 
corresponding to the sublimation temperature of graphites, and an
absorption feature centred at 9.7 \mus, long known to appear in
type 2 (obscured) AGN, attributed to silicate grains. 

The uncertain observational evidence for silicate in emission in type 1 AGN \citep{clavel00}, however,
posed a problem for the Unified Scheme according to which the various types of AGN can be explained 
by alignment effects
between the central sources, the obscuring material (torus) and the observer (\citealt{antonucci93}).
The problem was finally solved when silicates were unambiguously observed in emission in the MIR
spectra of many known AGN with IRS (\citealt{siebenmorgen05}; 
\citealt{sturm05}; \citealt{hao05}; \citealt{buchanan06}; \citealt{shi06}).

Since then, various studies of the behaviour of the silicates have appeared. The very first such works, based on a few tens of AGN \citep[e.g.][]{spoon07, hao07, wu09}, demonstrated that the silicate feature shows a wide diversity. On
average spectra it varies with AGN type, ranging from moderate emission in bright quasars to almost no emission or slight absorption in Seyfert 1 galaxies to stronger absorption in Seyfert 2 galaxies. Meanwhile, sparse reports on the detection 
of silicates in emission \citep{mason09, nikutta09} showed the diversity of the dust properties, albeit some of
them rather rare. 
Several of the above observations also revealed a second silicate feature at 18 \mus, as predicted by the models. 
The relative strength of the silicate features at 9.7 and 18 \mu has been put forward as a possible diagnostic
of the torus morphology \citep[e.g.][]{thompson09, feltre12} and chemistry \citep{sirocky08}. 

In this paper, we put together the largest sample of active galaxies with available IRS spectra ever composed (Sec. \ref{sec:sample}) with the aim to complement and extend previous studies on the MIR characteristics of AGN. 
To this aim, we apply a new spectral decomposition
technique to separate the nuclear emission from that of the host (Sec. \ref{sec:decomp}). We then proceed with a
thorough investigation of the behaviour of the silicate feature at 9.7 \mu and
18 \mu in the various AGN types (Sec. \ref{sec:silicates}).
Section \ref{sec:discuss} discusses our most important results and places
them into a more general context.

\section{The sample}\label{sec:sample}
 
The Cornell AtlaS of {\it Spitzer}/Infrared Spectrograph project (CASSIS\footnote{http://cassis.sirtf.com}; \citealt{lebouteiller11}) has made available the reduced spectra of all the sources observed with the low
resolution modules of IRS, a total of about 11000 unique observations. Our master sample is derived from 
the CASSIS version 6 catalogue, keeping each object that fulfills the following requirements: i) has an
identification in the NASA/IPAC Extragalactic Database\footnote{http://ned.ipac.caltech.edu} (NED); 
ii) has a robust (optical or infrared) 
spectroscopic redshift; iii) the IRS spectrum fully covers the range between 6 and 13 micron restframe; 
iv) the median signal-to-noise ratio (SNR) per pixel of the IRS spectrum is $>$2. 

To verify requirement i) we rely on the source cross-identification from CASSIS, which matches the source coordinates with the NED and SIMBAD databases \citep{lebouteiller11}.
For sources with no spectroscopic redshift in NED, we measure the redshift from the IRS spectrum using the template matching method presented in \citet{hernan12}. The typical redshift uncertainty with this method is 
$\Delta$$z$/(1+$z$) $\sim$ 0.002, well below the spectral resolution of our resampled spectra 
($\Delta$$\lambda$/$\lambda$=0.005--0.02). The resampling also increases the minimum SNR per resolution element from 2 to $>3$ (see Sec. \ref{sec:decomp}). 

Taking these criteria into account and removing duplicate entries, we end up with a list of 2299 extragalactic objects. 
Out of these, 784 objects have a NED classification as AGN of types 1, 2 or intermediate, and this is the sample we will be working with henceforth. AGN of undefined type were left out of the sample. 
Among the type 1 AGN lie 141 quasars from the Sloan Digital Sky Survey (SDSS) 
Data Release 7 Quasar Catalogue \citep{shen11}, that for few specific purposes will be examined separately.
The numbers of the object per subsample are shown in Table \ref{tab:samples}. The AGN sample is 
heterogeneous but, nevertheless, representative of the infrared (IR) AGN population.

\begin{table}
\begin{center}
\caption{Number of objects per AGN type. }
\label{tab:samples}
\begin{tabular}{llll}
\hline
 Type & N$_{\rm obj}$ & Type & N$_{\rm obj}$ \\
\hline
Type 1 AGN         & 363   & Sy1.2  & 24 \\
Type 2 AGN         & 325   & Sy1.5  & 32 \\
                               &           & Sy1.8  & 18 \\
SDSS quasars    & 141   & Sy1.9  & 22 \\
\hline
\end{tabular}
\end{center}
\end{table}

\section{Spectral decomposition}\label{sec:decomp}

The observed MIR spectra of AGN are affected by the presence of their host galaxies in two important ways. 
One is the absorption or scattering of AGN emissions by material (gas and dust) in the AGN 
line of sight. This so-called foreground absorption modulates the AGN spectrum with a 
multiplicative factor $e^{-\tau(\lambda)}$, where $\tau(\lambda)$ represents the optical 
depth at wavelength $\lambda$. It is not possible to distinguish foreground absorption from intrinsic AGN 
absorption (that is, the one produced in the AGN torus) from MIR data alone, 
since similar extinction laws are considered to apply to dust grains in the torus and the host. 
The amount of foreground extinction varies from source to source, but at MIR wavelengths it 
is expected to be mild in most sources, with the exception of some dusty starbursts and 
edge-on spiral galaxies. The other important effect on the AGN spectra is the contamination 
from host galaxy emission that blends with the AGN spectrum. The importance of this 
background emission depends on the relative luminosities of the AGN and the host 
and --crucially-- on the spatial resolution of the spectroscopic observations. Since the emission 
from the AGN is typically unresolved, an increase in the spatial resolution implies that a larger fraction 
of the host emission can be resolved away. In any case, the background emission represents 
an additive modification to the AGN spectrum.

The purpose of our spectral decomposition is to separate the AGN and host emissions in the 
integrated AGN+host spectra. If successful, this decomposition allows to study the AGN 
emission as if the host galaxy was resolved away. To this aim we employ the decomposition 
method presented in \cite{hernan15}. The method relies on the large number of high-quality 
spectra in CASSIS to reproduce the spectra of composite sources as a linear combination of three 
CASSIS spectra, each selected from subsamples of sources whose mid-IR emission is completely 
dominated by the AGN, the star-formation, or the stellar population. We select these 
`single-spectral-component' templates as follows:

\begin{figure*}
\begin{center}
\includegraphics[width=17cm]{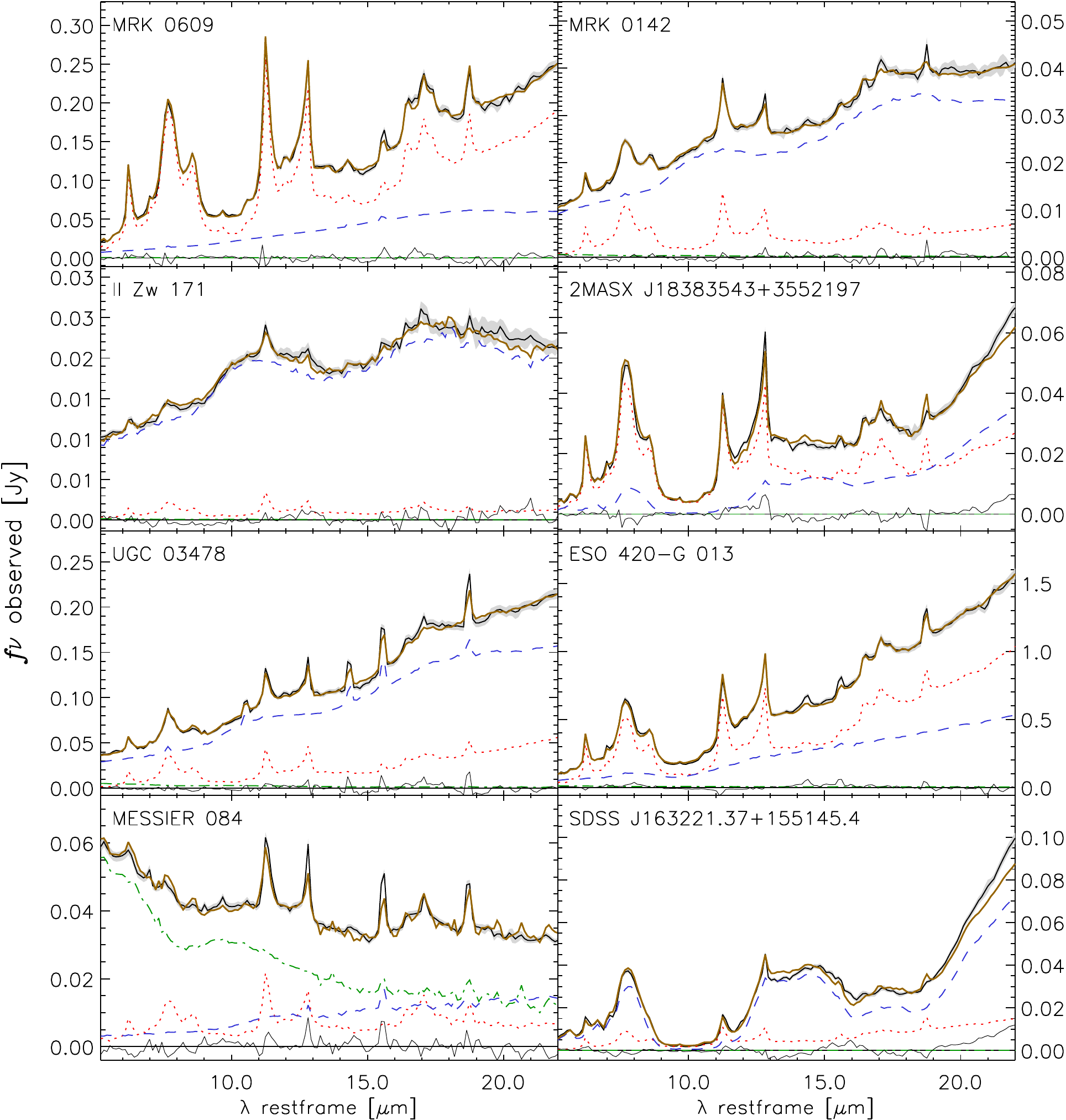}
\caption{Examples of best-fitting decomposition models for the IRS spectra. 
The black solid line with grey shading represents the IRS spectrum (resampled at $\Delta\lambda$=0.1 \mu 
resolution) and its 1-$\sigma$ uncertainty (photometric errors only). The dotted, dashed, and dot-dashed lines 
represent, respectively, the PAH, AGN, and stellar components of the best-fitting model, shown in yellow. 
The thin solid line at the bottom of each plot represents the residual (spectrum - model).}
\label{fig:decompositions}
\end{center}
\end{figure*}

For the `stellar' templates, we select 19 local elliptical and S0 galaxies. To ensure they have negligible star 
formation, we require the Polycyclic Aromatic Hydrocarbons (PAH) bands to be very weak or absent, with 
equivalent widths for the 6.2 \mu (EW$_{62}$) and 11.3 \mu (EW$_{113}$) PAH features $<$0.02 \mus. We also 
check that the IRS spectra have a blue stellar-like MIR continuum and the sources are not classified as AGN 
in NED. The 54 star-forming templates (`PAH' templates) are IRS spectra of normal star-forming and 
starburst galaxies at redshifts up to $z$=0.14. We make sure these sources do not have significant 
stellar contributions to their MIR spectra by requiring both high EW of the PAH features (EW$_{62}$ $>$ 
1.0 \mu and EW$_{113}$ $>$ 1.0 \mus) and a very weak continuum at 5 \mus. We also verify that they are not 
classified as AGN in NED. Finally, the 147 `AGN' templates are IRS spectra of sources classified in the 
optical as quasars, Seyfert galaxies,  LINERs, and blazars. We also include a variety of optically obscured 
AGN and radiogalaxies. The templates include sources at redshifts from $z$=0.002 to $z$=1.4 and cover 
several orders of magnitude in bolometric luminosity. We ensure that the AGN templates do not contain 
any significant emission from the host galaxy by requiring the PAH features to be extremely weak or 
absent (EW$_{62} <$ 0.02 \mu and EW$_{113} <$ 0.02 \mus). 

Because the AGN and PAH templates are real spectra, each of them already includes some amount 
of foreground extinction built in. Therefore, we rely on the large number of AGN and PAH templates to 
reproduce the diversity of observed spectra that arises from different levels of foreground extinction as 
well as source to source variation in the intrinsic AGN and host spectra. This approach has the advantage
of not depending on assumptions about the --unobservable-- intrinsic AGN spectrum or the extinction law. 
Obtaining a good fit with this decomposition method requires finding an AGN template with the appropriate 
level of foreground absorption. This can be problematic for sources with very deep absorption features, 
since few pure-AGN spectra have them. Accordingly, decomposition results for sources with deep 
absorption features have larger residuals and uncertainties.

We separate the AGN sample into two groups depending on the spectral coverage:
those objects with silicate features observed both at 9.7 and at 18 \mu in their full extent 
and those for which only the feature at 9.7 \mu is covered. 
For sources in the first group, we fit the spectral range between 5.2 and 22 \mu restframe, while 
for those in the second, we fit only the 5.2 to 15.8 \mu range. We resample both the spectra and 
templates to a common wavelength grid with an uniform wavelength resolution of $\Delta\lambda$=0.1 \mus. 
This increases the SNR per resolution element by 
$\sim$60\% on average, while it still allows to resolve important features such as the PAH bands.  
For every galaxy in the sample we try spectral decompositions using every possible combination of a stellar 
template, a PAH template, and an AGN template. The best fitting model is the one that produces the absolute 
minimum of $\chi^2.$ However, to calculate expected values for observables (e.g. the luminosity of 
the AGN component or the strength of the silicate feature) and their uncertainties, we use the full probability 
distribution functions (PDFs) calculated with the `max' method described in \citet{noll09} \citep[for details see][Sec. 2]
{hernan15}. The method also yields, for each object, the fractional contribution of each the 
three components to the total luminosity, in the wavelength range of interest.

Thanks to the use of large sets of real spectra as templates, our decomposition method manages to reproduce the MIR spectrum of composite sources with unprecedented accuracy (see Fig. \ref{fig:decompositions}). 
Typical $\chi^2$ values are lower than 2, indicating that residuals in the model fits are dominated by noise in both the 
spectra and templates for most sources.

\section{The Silicate features}\label{sec:silicates} 

The silicate features at 9.7 and 18 \mu observed in the IR
spectra of AGN are believed to arise from the inner, hotter parts of the torus or the hot, illuminated side of the 
clumps. We define the strength of the silicate feature following \cite{pier92}:

\begin{equation}
{\rm S_{\lambda}}= {\rm ln} \frac{{\rm F}(\lambda_{peak})}{{\rm F}_{\rm c}(\lambda_{peak})}
\label{eq:ssil}
\end{equation}

\noindent
where F($\lambda_{\rm peak}$) and F$_{\rm c}(\lambda_{\rm peak})$ are the flux densities of the 
spectrum and the underlying continuum at the peak wavelength of the features, ${\lambda_{\rm peak}}$. 
A negative (positive) value indicates a feature in absorption (emission).

\subsection{The Silicate feature at 9.7 micron}\label{sec:silicates97}

\begin{figure}
\begin{center}
\includegraphics[width=8cm,angle=0]{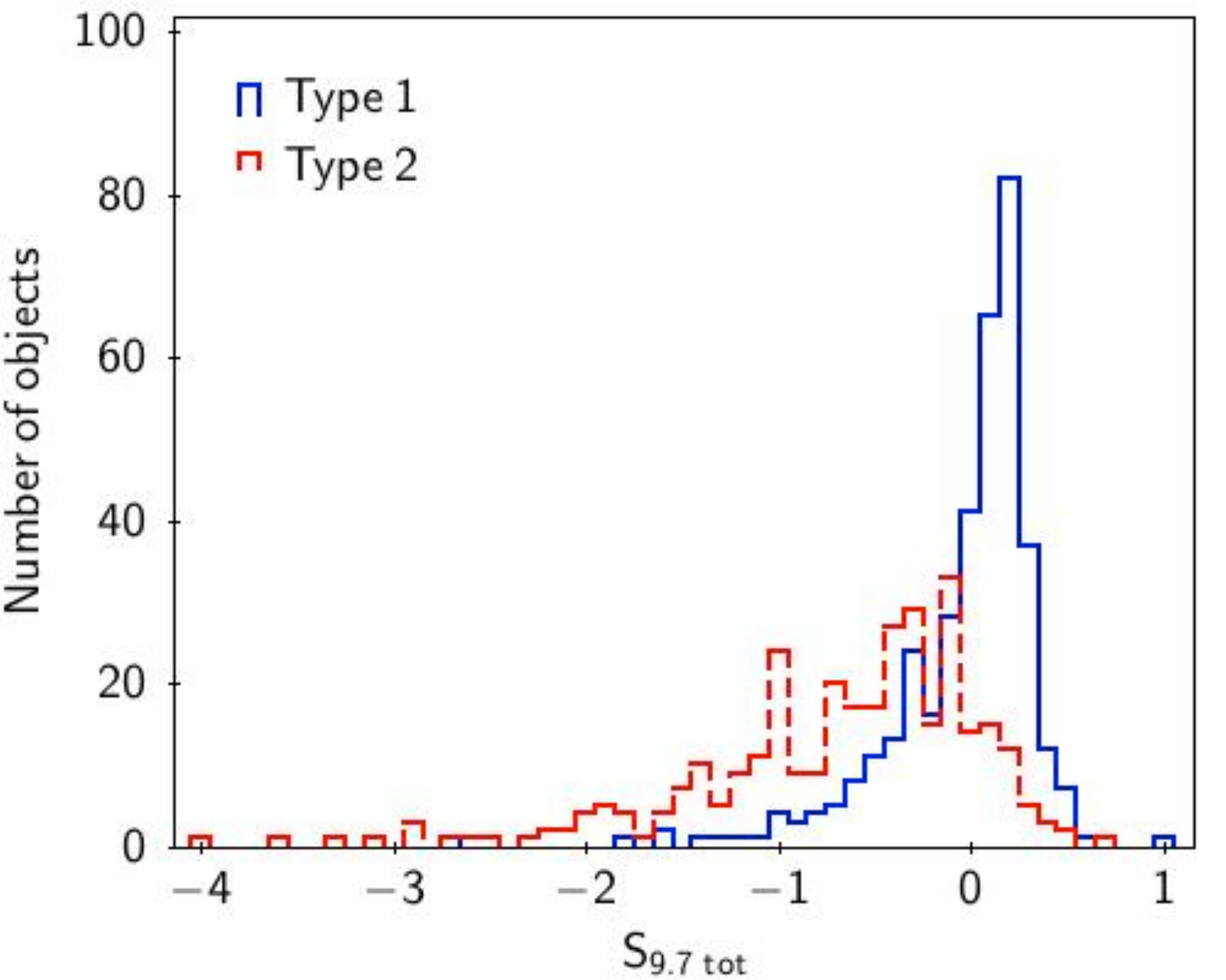}
\includegraphics[width=8cm,angle=0]{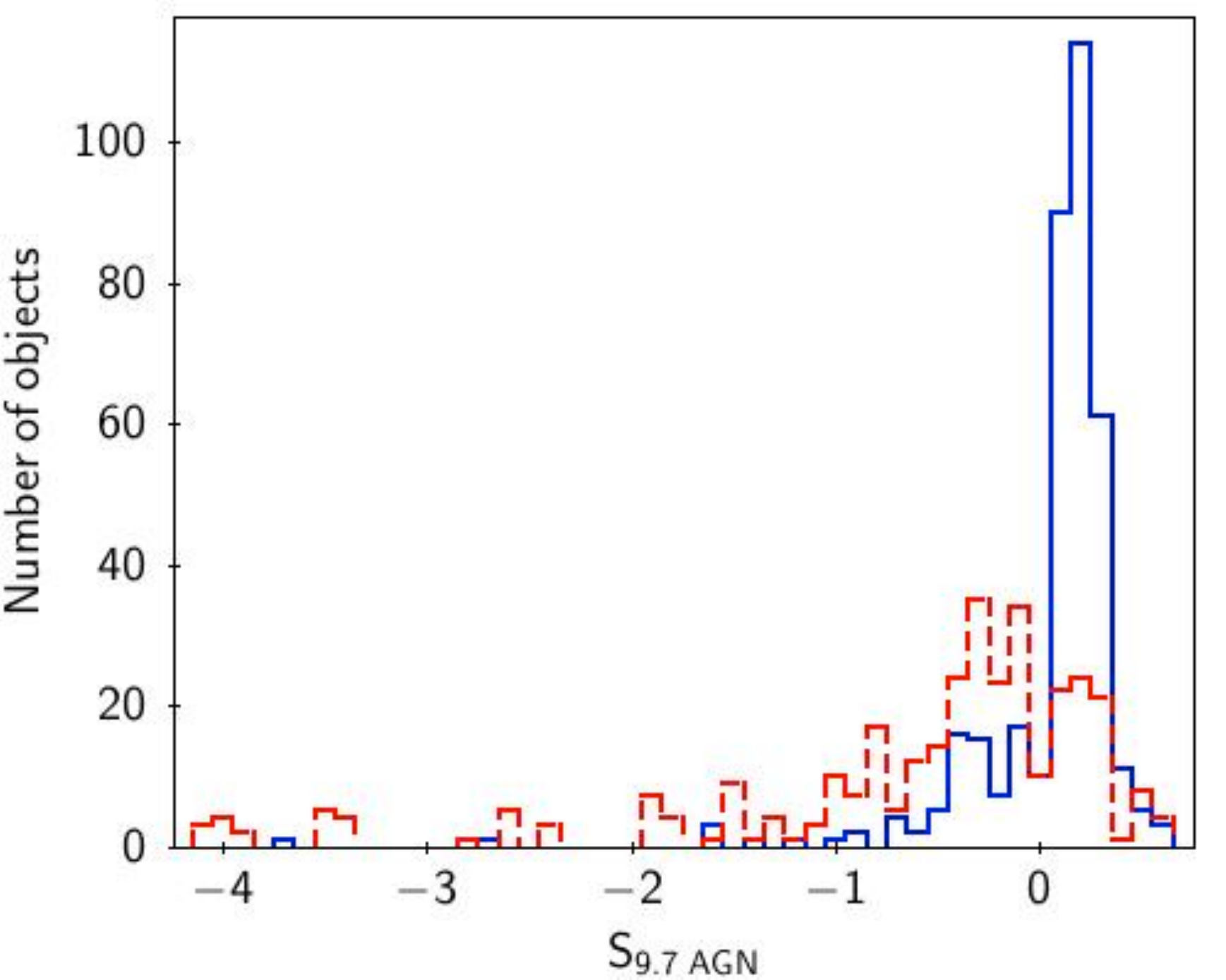}\\
\caption{S$_{\rm 9.7}$ distribution per for type 1 (blue solid histogram) and type 2 (red dashed histogram),
before (top panel) and after (bottom panel) the subtraction of the host galaxy.}
\label{fig:ssilirshisto}
\end{center}
\end{figure}

The top panel of Fig. \ref{fig:ssilirshisto} shows the distribution of the strength
of the silicate feature at 9.7 \mu as measured on the original IRS spectra, S$_{\rm 9.7 \,\, tot}$, 
for type 1 and type 2 AGN (the 96 AGN of intermediate type are not included here and will be discussed separately).

As already observationally established in the past ten years, S$_{\rm 9.7 \,\, tot}$
takes a wide range of values \citep[e.g.][just to name a few]{hao07, 
levenson07, spoon07, wu09}. The average spectra of type 1 AGN exhibit the feature in weak 
to moderate emission  \citep[see e.g.][]{hao07,wu09}, while those of type 2 AGN present the 
feature in absorption \citep[][]{sturm06, schweitzer08, mason09, hernan11}. 
Individually, however, type 1 and type 2 AGN can show silicates in absorption and emission, 
respectively. In our sample of 698 type 1 and 2 AGN, 35\% of the type 1 AGN present the 
feature in absorption and about 15\% of the type 2 AGN show the feature in emission. 

\begin{figure}
\begin{center}
\includegraphics[width=8.5cm,angle=0]{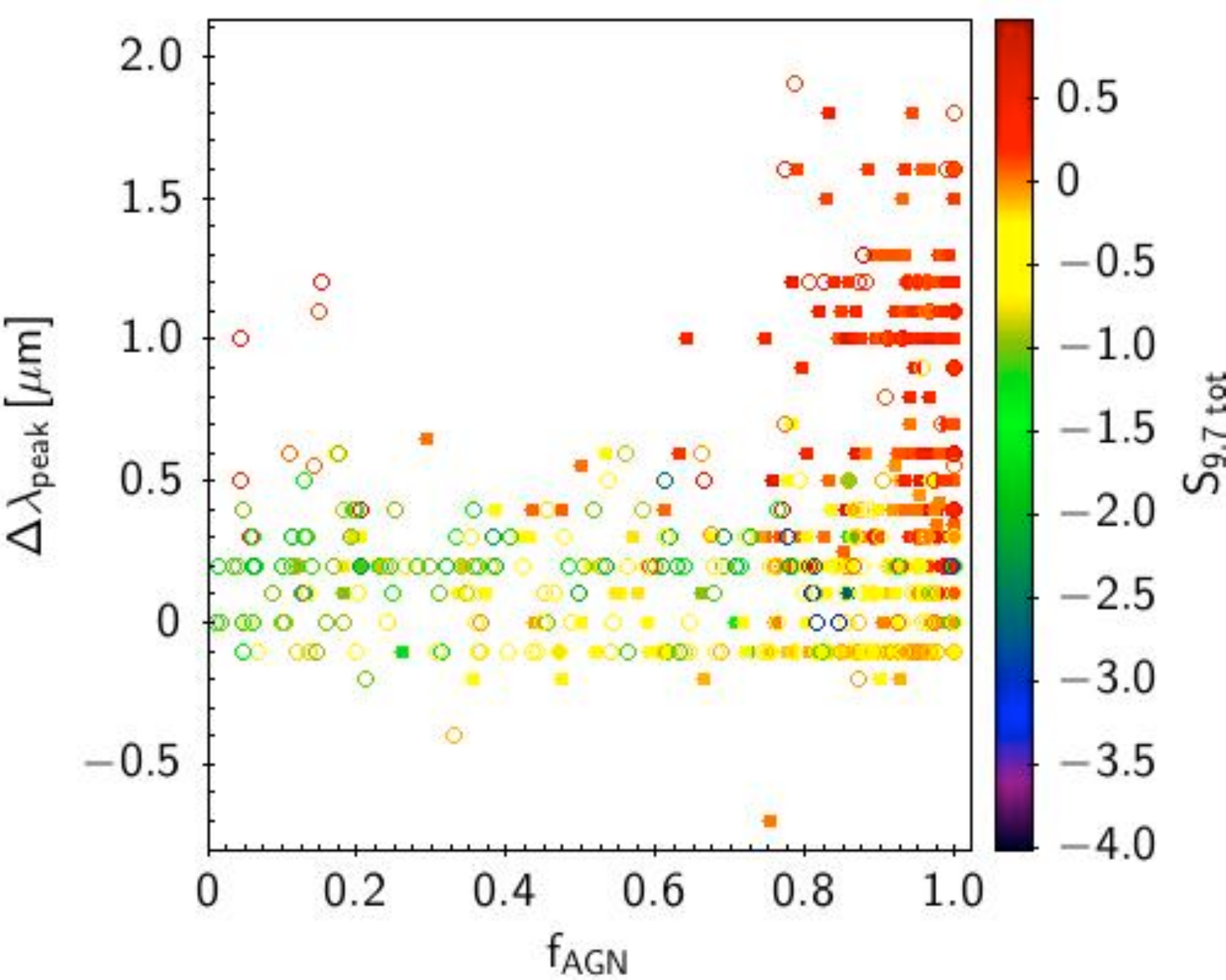}\\
\caption{${\rm \Delta \lambda}_{\rm peak}$ as a function of $f_{\rm AGN}$, for type 1 and type 2 objects (filled and open
symbols, respectively). The symbols are colour-coded based on the value of S$_{\rm 9.7 \,\, tot}$. 
The quantisation of ${\rm \Delta \lambda}_{\rm peak}$ is an artifact of the algorithm that measures 
$\lambda_{\rm peak}$.}
\label{fig:lpeakssil}
\end{center}
\end{figure}

At the same time, when in emission, the peak of the feature is often 
shifted to wavelengths longer than 9.7 \mu in the rest frame, while the shift
affects much less the feature when in absorption, as already reported by e.g \cite{shi14}. 
Fig. \ref{fig:lpeakssil} shows the distribution of the shift,
$\Delta\lambda_{\rm peak} = \lambda_{\rm peak}-9.7$ \mus, as a function of the fractional 
contribution of the AGN to the luminosity in the range between 5 and 15 \mus, $f_{\rm AGN}$,
for type 1 and type 2 objects (filled and open symbols, respectively), colour-coded by S$_{\rm 9.7 \,\, tot}$. 
Looking at the sample as a whole, 65\% (20\%) of the objects with the silicates in emission 
have their $\lambda_{\rm peak} >$ 10.2 \mu ($\lambda_{\rm peak} > 10.6$ \mus), 
while the fraction of objects with the same amount of shift among the AGN with silicates in 
absorption is less than 3\%. 
The shift to longer wavelenghts is largely associated to a silicate feature in emission, and this in turn 
only occurs in strongly AGN-dominated spectra.

\subsection{Removing the effects of the host}\label{sec:silagn}

\begin{figure}
\begin{center}
\includegraphics[width=9cm,angle=0]{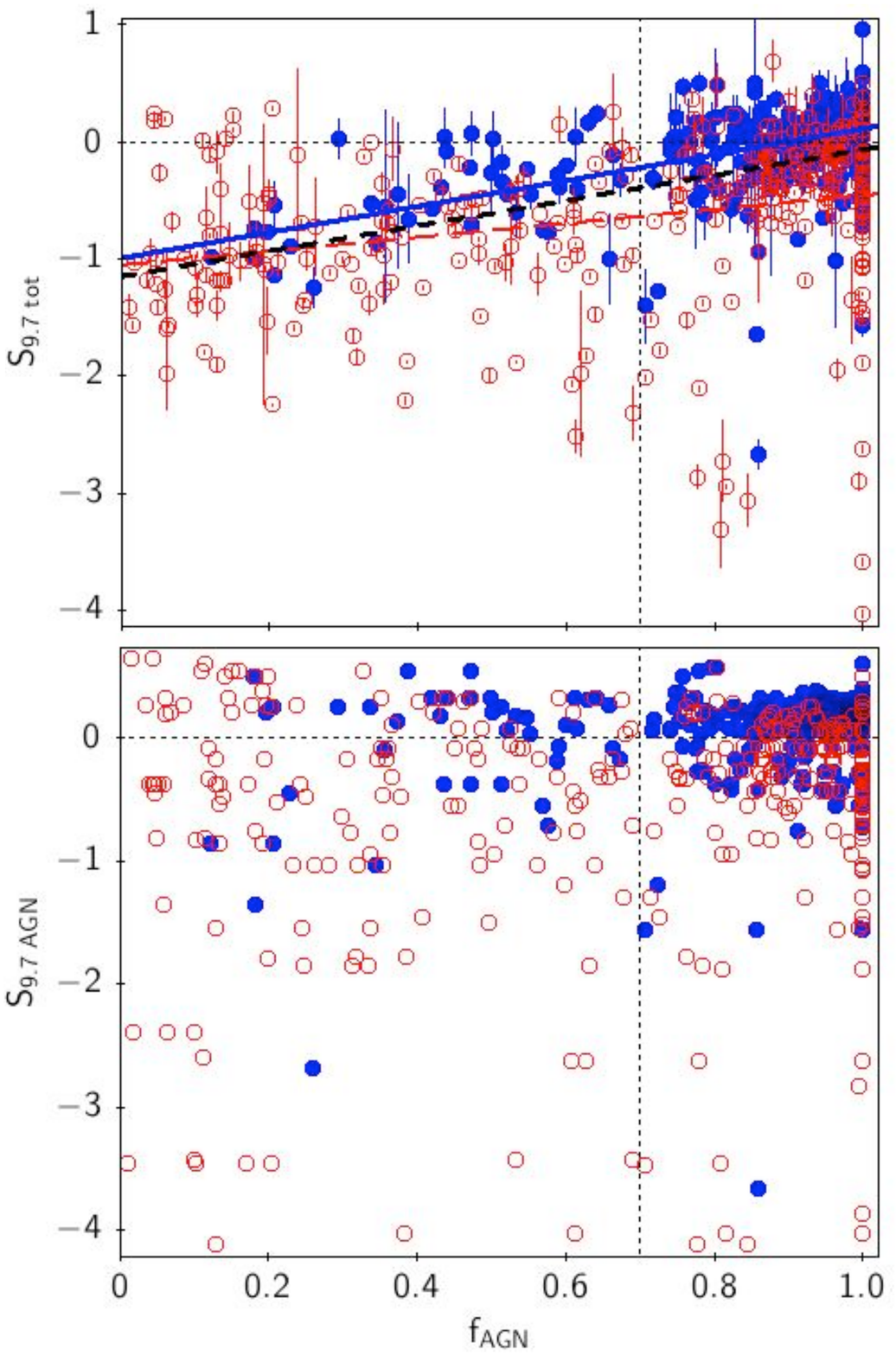}\\
\vskip -1.3cm
\caption{S$_{\rm 9.7}$ as a function of $f_{\rm AGN}$,
before (upper panel) and after (lower panel) the subtraction of the host galaxy. Filled and open
circles denote type 1 and type 2 AGN, respectively. The thin dashed lines show S$_{\rm 9.7}$=0.0
and $f_{\rm AGN}$=0.7, the thick dashed line shows the weak ($r$=0.45) linear correlation of the full
sample, and the solid and long-dashed lines show the correlations for the type 1 ($r$=0.51) and
type 2 ($r$=0.26) objects, respectively.}
\label{fig:ssilfagn}
\end{center}
\end{figure}

As S$_{\rm 9.7 \,\, tot}$ are measured on the original IRS spectra, we expect the derived values to be contaminated 
by the emission of the host galaxy for all objects but those for which the AGN completely dominates the MIR emission.
The top panel of Fig. \ref{fig:ssilfagn} shows S$_{\rm 9.7 \,\, tot}$ as a function of $f_{\rm AGN}$.
The filled and open symbols correspond to type 1 and type 2 AGN, respectively. Error bars for S$_{\rm 9.7 \,\, tot}$ 
are shown in this plot but they will not be repeated in following figures, 
in order to keep the plots as little crowded as possible. What we see here is that as the contribution
of the host becomes more important (i.e. as $f_{\rm AGN}$ decreases) the strength of the silicate feature decreases, 
with only few objects exhibiting S$_{\rm 9.7 \,\, tot} > 0.0$ for $f_{\rm AGN} < 0.7$. 
The dashed line shows the (weak) trend for the full sample, with a linear correlation coefficient of $r$=0.45.
The trend, however, is driven by type 1 objects (filled symbols and corresponding solid line) 
due to the contamination by the emission of the host galaxy.

In order to see how the silicates behave in the vicinity of the nucleus, we need 
to remove the contribution of the host galaxy, applying the spectral decomposition procedure described in Sec.
\ref{sec:decomp}. The distribution of the strength of the silicate feature on the host-subtracted spectrum, 
S$_{\rm 9.7 \,\, AGN}$, is shown in the bottom panel of Fig. \ref{fig:ssilirshisto}. The behaviour of 
S$_{\rm 9.7 \,\, AGN}$ with $f_{\rm AGN}$ shown in the lower panel of Fig. \ref{fig:ssilfagn} differs from that of 
S$_{\rm 9.7 \,\, tot}$ in that there is now no correlation between the two quantities, confirming that the correlation
was due to the contamination from the emission of the host, indeed.

\begin{figure}
\begin{center}
\includegraphics[width=8cm,angle=0]{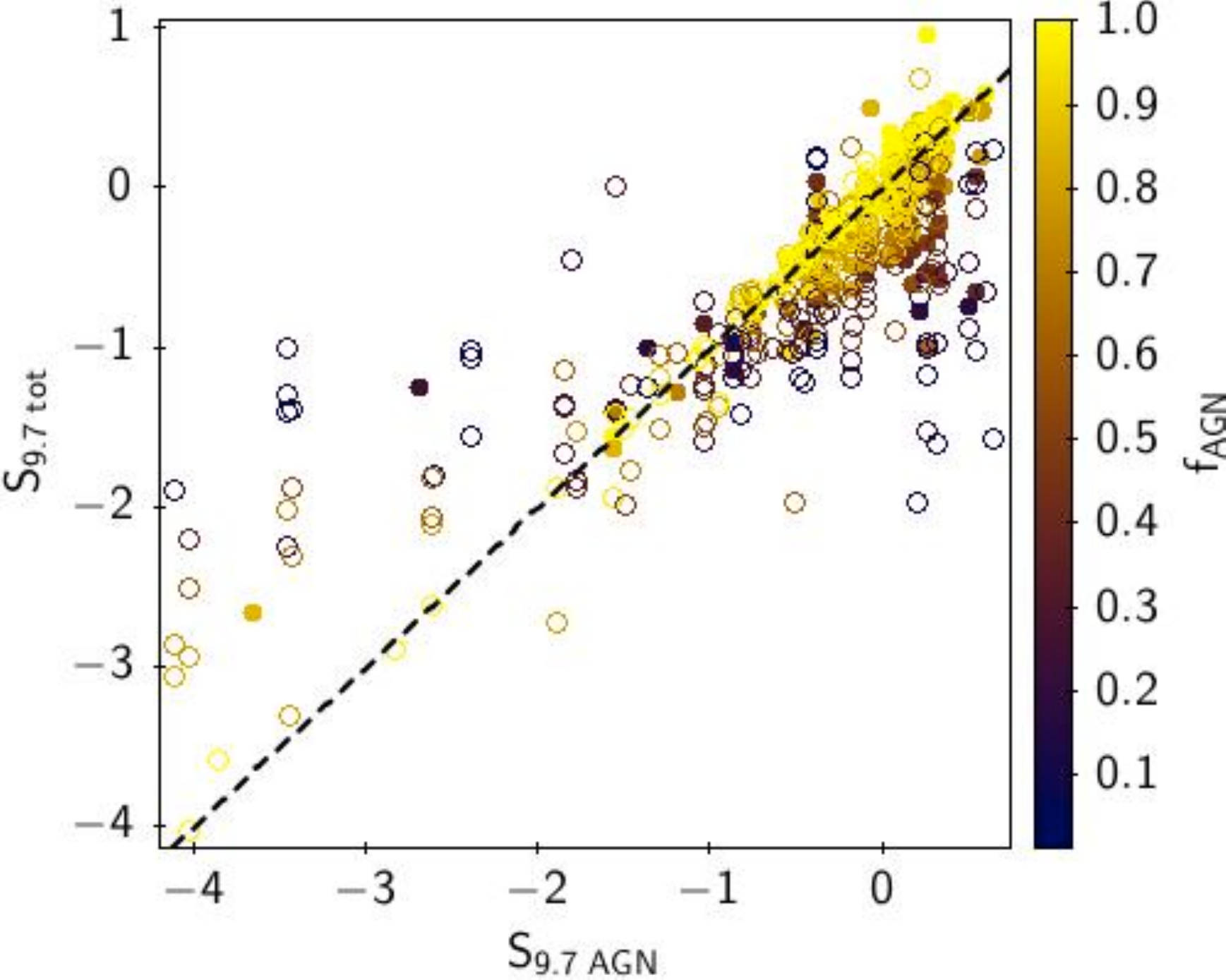}\\
\caption{S$_{\rm 9.7 \,\, tot}$ versus S$_{\rm 9.7 \,\, AGN}$, colour-coded based on$f_{\rm AGN}$. 
Filled and open symbols denote type 1 and type 2 AGN, respectively.}
\label{fig:silvssil}
\end{center}
\end{figure}

A direct comparison of the two measurements of the silicate feature at 9.7 \mu is shown in Fig. \ref{fig:silvssil}, 
colour-coded by the value of $f_{\rm AGN}$. Objects with MIR emission completely dominated by the AGN 
(light-colour symbols) are not affected by the subtraction of the host (they lie on or very near the 1:1 line). 
However, as the contribution of the host galaxy becomes more important, i.e. $f_{\rm AGN}$ decreases, 
(symbols become darker in Fig. \ref{fig:silvssil}), the points deviate more and more from the 1:1 line. 

By subtracting the emission of the galaxy, the number of type 1 AGN with silicates in emission increases by 20\%,
reaching 80\% of all type 1 AGN, while the number of type 2 AGN with the feature
in emission doubles, reaching a total of 25\%. At the same time, 35\% of both type 1 and type 2 AGN
with the feature in absorption exhibit the feature in even deeper absorption once the emission from
the host is removed. This happens because AGN with a silicate feature in deeper absorption than that 
of the surrounding host get their silicate feature `refilled' in the integrated spectrum.

\begin{figure}
\begin{center}
\includegraphics[width=8.5cm,angle=0]{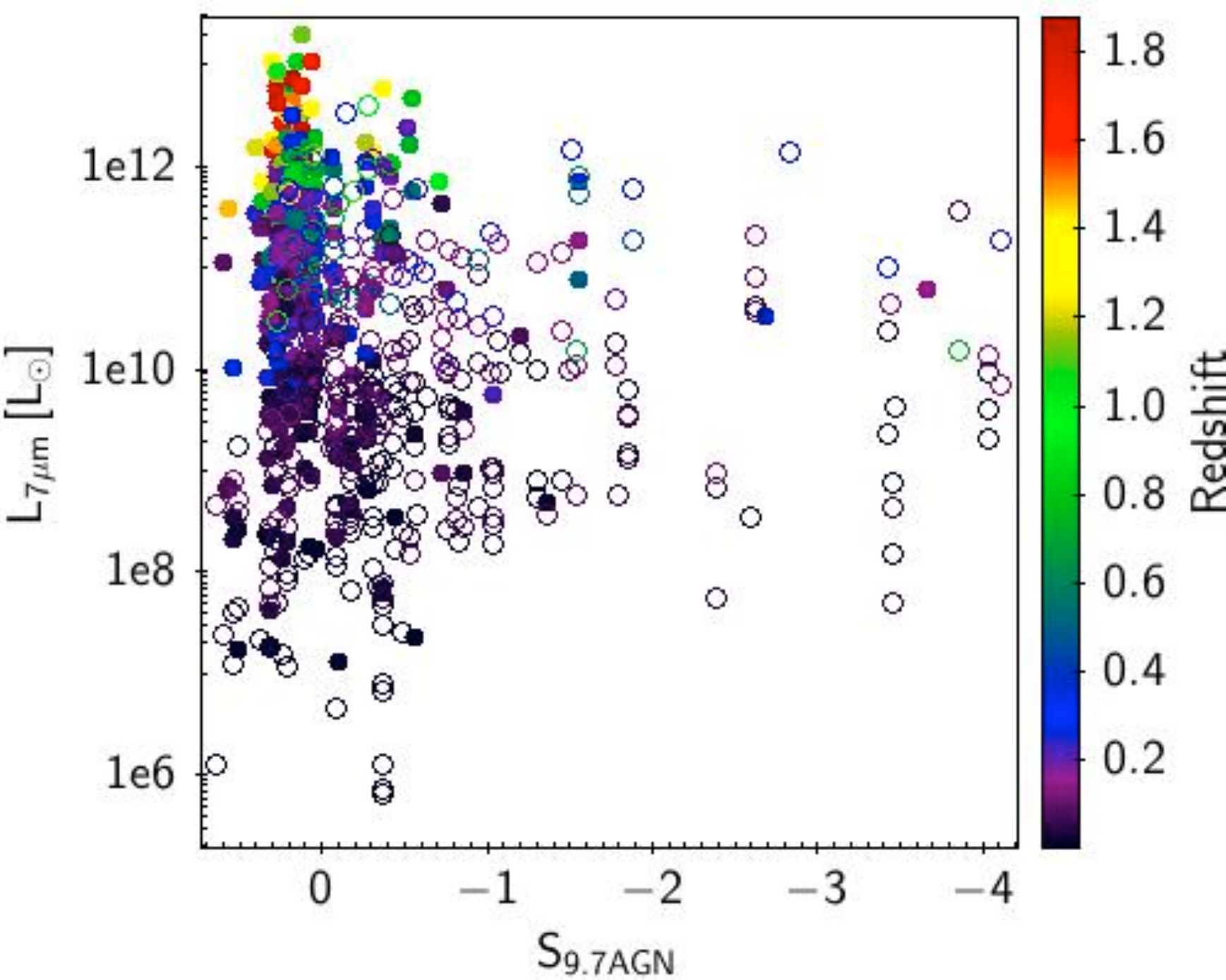}
\caption{The relation between S$_{\rm 9.7 \,\,AGN}$ and L$_{\rm 7}$, with points colour-coded based on their
redshift. The filled and open circles correspond to type 1 and type 2 AGN, respectively.} 
\label{fig:ssill7}
\end{center}
\end{figure}

In order to check whether S$_{\rm 9.7 \,\, AGN}$ is affected by the AGN luminosity, as
proposed by e.g. \cite{maiolino07}, we have to rely on the luminosity at 7 \mus, L$_7$, that spans
over six orders of magnitude in our sample.
MIR luminosity in AGN has, in fact, been shown to tightly correlate with the X-ray 
luminosity \citep[see e.g.][]{lutz04,horst08,mateos15}, the ratio of the MIR to the bolometric luminosity
depends, however, on the column density along the line of sight \citep[Fig. 21 in ][]{hatzimi09}.
Fig. \ref{fig:ssill7} shows S$_{\rm 9.7 \,\, AGN}$ as a function of L$_{\rm 7}$ measured on the
galaxy-subtracted spectrum,
with the points coloured based on the redshift. There is clearly no
dependence of S$_{\rm 9.7 \,\, AGN}$ with L$_{\rm 7}$
and the features can be in emission (S$_{\rm 9.7}>0.0$) even in the faintest AGN.
Deep silicate features (S$_{\rm 9.7} < $-2) are only found at intermediate luminosities (or redshifts). 
A deeply obscured but low-luminosity AGN would be 
overwhelmed by the emission of its host, that would `refill' the silicate feature. The lack of deep 
silicates in objects with very high IR luminosities and/or high redshifts, on the other hand, suggests a selection effect,
as they might be too faint in the optical for a reliable identification.

\subsubsection{AGN of intermediate types}\label{sec:interm}

\begin{figure}
\begin{center}
\includegraphics[width=8cm,angle=0]{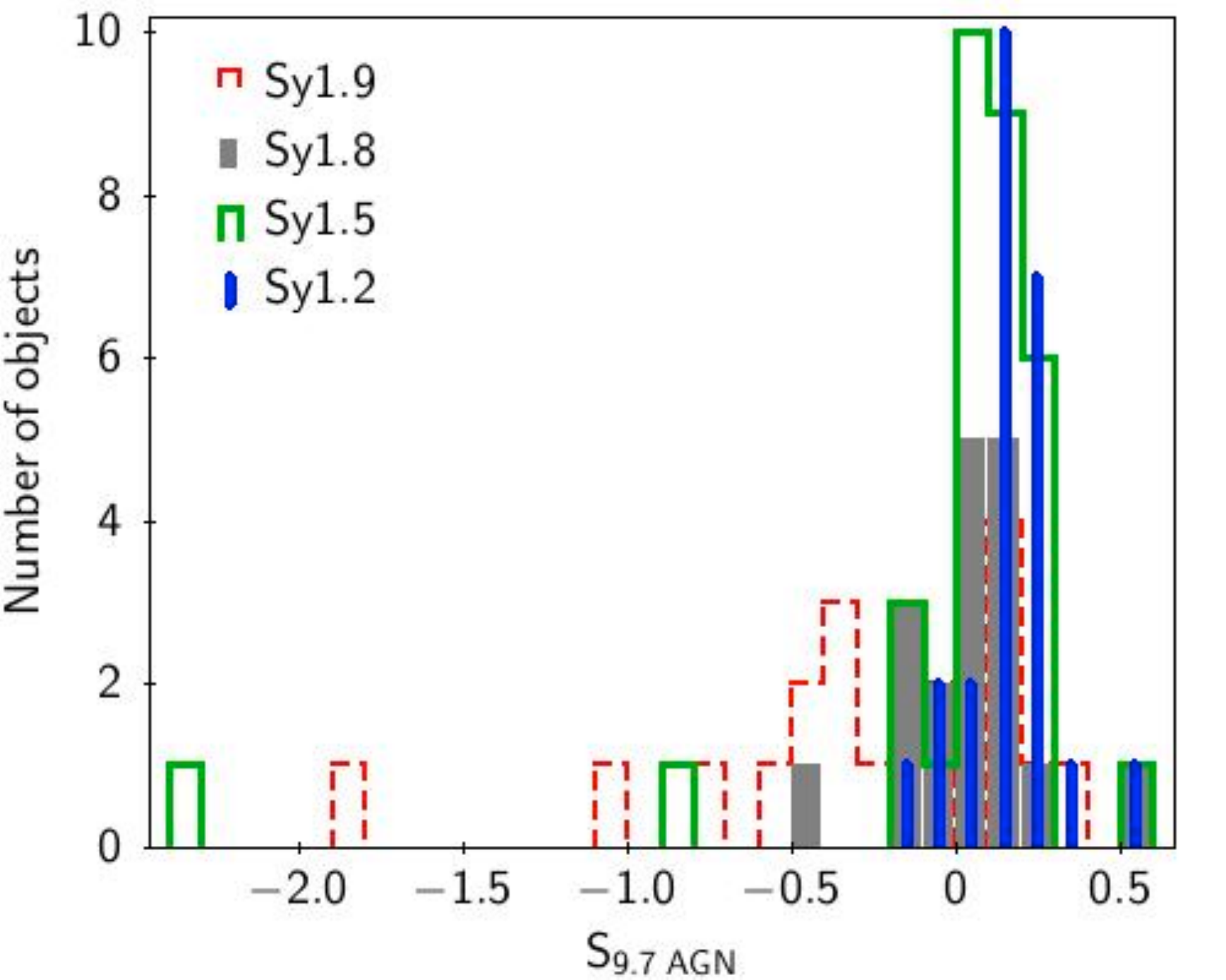}
\caption{The distribution of S$_{\rm 9.7 \,\, AGN}$ for the four intermediate types of Seyfert galaxies.}
\label{fig:ssilinterm}
\end{center}
\end{figure}

Among the 784 AGN of our sample, 96 have a NED classification of intermediate type Seyfert galaxies.
The number of objects per type (comparable in all four sub-samples) is shown in Table \ref{tab:samples}. 
Figure \ref{fig:ssilinterm} shows the distribution of S$_{\rm 9.7 \,\, AGN}$ for the four sub-samples.

The mean average S$_{\rm 9.7 \,\,AGN}$  decreases from Sy1.2 towards later types, with Sy 1.2, 1.5 and 1.8 
showing on average the silicate feature in very weak or no emission, while Sy1.9  
has a negative average value. 
Additionally, as we move to later intermediate types, the dominance of the AGN component
to the MIR emission decreases: while 96\% of Sy1.2 objects have $f_{\rm AGN}>0.7$, the fraction drops to 55\% for
Sy1.9s. This is in agreement with \cite{deo07} that found the MIR spectra of
Sy1.8 and Sy1.9 to be dominated by starburst features (PAH).
The mean values of S$_{\rm 9.7 \,\, AGN}$ and the fraction of objects with $f_{\rm AGN}>0.7$ for
each intermediate type are shown in Table \ref{tab:interm}. Note that the shift towards lower values of
S$_{\rm 9.7 \,\, AGN}$ when moving towards later Seyfert types persist even when only objects with $f_{\rm AGN}>0.7$
are considered, as seen in the right-most column of table \ref{tab:interm}. Finally, the $\lambda_{\rm peak}$
of intermediate Seyfert types shows the same behaviour as for the rest of the AGN sample, i.e. as described in
Sec. \ref{sec:silicates97}.

\begin{table}
\begin{center}
\caption{Mean values and standard deviations of S$_{\rm 9.7}$ for each intermediate AGN type, 
fraction of objects with $f_{\rm AGN} > 0.7$ and S$_{\rm 9.7}$ for that fraction.}
\label{tab:interm}
\begin{tabular}{lrcr}
\hline
 Type & $\langle$S$_{\rm 9.7 \,\, tot} \rangle$ & $f_{\rm AGN}>0.7$ & $\langle$S$^{f_{\rm AGN}>0.7}_{\rm 9.7 \,\, tot}\rangle$\\
\hline
Sy1.2       & 0.175$\pm$0.13  & 96\% & 0.160$\pm$0.14\\
Sy1.5       & 0.043$\pm$0.46  & 90\% & 0.087$\pm$0.22\\
Sy1.8       & 0.022$\pm$0.22  & 61\% & 0.059$\pm$0.11\\
Sy1.9       & -0.305$\pm$0.49 & 55\% & -0.227$\pm$0.35\\
\hline
\end{tabular}
\end{center}
\end{table}

\subsubsection{SDSS quasars}\label{sec:sdss}

\begin{figure}
\begin{center}
\includegraphics[width=8.5cm,angle=0]{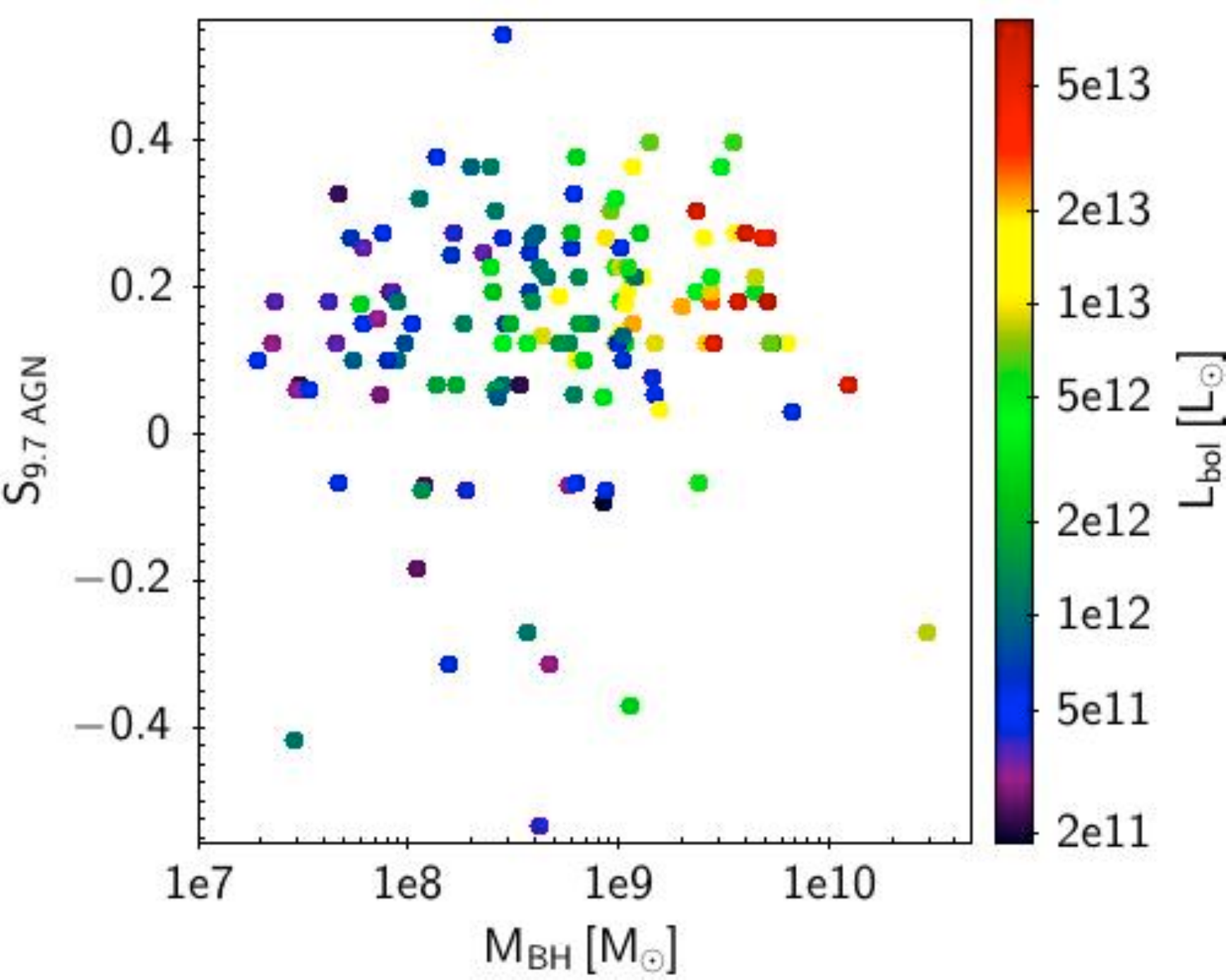}
\caption{S$_{\rm 9.7 \,\, AGN}$ as a function of M$_{\rm BH}$ for the 141 SDSS quasars of the sample.
The points are colour-coded by L$_{\rm bol}$.}
\label{fig:ssilmbh}
\end{center}
\end{figure}

Out of the 784 AGN, 141 are spectroscopically confirmed SDSS quasars, for which
estimates of the mass of the central black hole, M$_{\rm BH}$, derived from emission line 
measurements, as well as bolometric luminosities, L$_{\rm bol}$, derived from fitting techniques 
are available \citep{shen11}. \cite{maiolino07}
reported an increase of S$_{\rm 9.7}$ with increasing M$_{\rm BH}$, from low-luminosity, low-redshift type 1 AGN to
high-luminosity, high-redshift quasars. \cite{thompson09}, on the other hand, found no trend of with luminosity. 
M$_{\rm BH}$ for the 141 SDSS quasars in question spans the range [10$^{7.3}$ M$_{\odot} - 10^{10}$ M$_{\odot}$], i.e. almost identical
to that of the \cite{maiolino07} sample, but we do not find any correlation of S$_{\rm 9.7 \,\, AGN}$ with M$_{\rm BH}$, 
as shown in Fig. \ref{fig:ssilmbh}.
Note, however, that  the bolometric luminosities of the SDSS quasars of our sample are all above 8.2 $\times 10^{10}$
L$_{\odot}$ (10$^{44.5}$erg/sec), i.e. the two samples are not directly comparable. In the quasar
sub-sample, S$_{\rm 9.7 \,\, AGN}$ and L$_{\rm bol}$ are completely uncorrelated (see colour-coding in
Fig. \ref{fig:ssilmbh}), in agreement with \cite{thompson09}.

\subsection{The Silicate feature at 18 micron}\label{sec:silicates18}

\begin{figure}
\begin{center}
\includegraphics[width=8.5cm,angle=0]{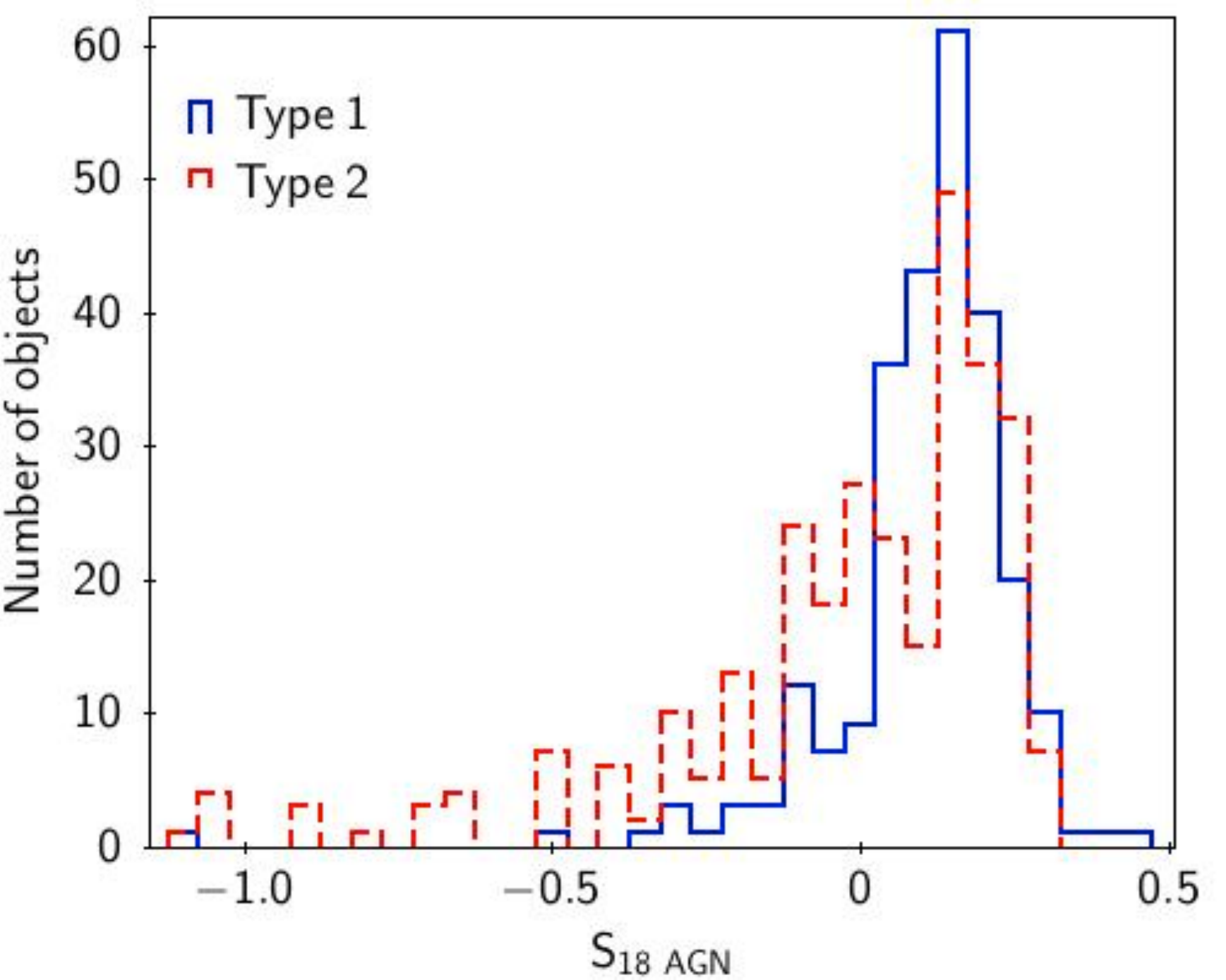}
\caption{S$_{\rm 18 \,\, AGN}$ distribution per AGN type (type 1 in blue solid histogram, type 2 
in red dashed histogram).}
\label{fig:s18}
\end{center}
\end{figure}

The silicate feature at 18 \mus, though predicted by models and observed in many of the low-to-intermediate
redshift AGN ($z$ typically $\le 0.5$) by now \citep[e.g.][]{hao05,sirocky08,thompson09}, 
has received less attention than its lower wavelength counterpart, as its measurement presents a 
greater challenge. For one, it often overlaps with the MIR bump of the AGN SED \citep[e.g.][]{prieto10}. 
Also, the steep mid-to-far IR emission from the host implies it contributes a higher fraction to the continuum 
emission at these longer wavelengths, making the measurement of the feature
a tedious job. Following the procedure described in Sec. \ref{sec:decomp}, but extending the wavelength range to
22 \mu restframe, we perform spectral decompositions and measure the strength of the 18 \mu silicate 
feature in the AGN component, S$_{\rm 18 \,\, AGN}$, for the 631 AGN with adequate
wavelength coverage. Fig. \ref{fig:s18} shows the distribution of S$_{\rm 18 \,\, AGN}$ for type 1 (blue
solid histograms) and type 2 (red dashed histograms). 

While at least as prominent as its
counterpart at 9.7 \mu in emission, when in absorption the feature only reaches moderate depths.
Furthermore, more than 50\% of type 2 AGN exhibit the feature in emission while only
10\% of type 1 AGN have it in absorption. Comparing these numbers with those for 
S$_{\rm 9.7 \,\, AGN}$ it becomes obvious that the feature can be in absorption at 9.7 \mu while
still in emission at 18 \mus. We will get back to this point in Sec. \ref{sec:discuss}.

\section{Discussion and Conclusions}\label{sec:discuss}

Using the largest sample of AGN with MIR spectroscopy ever assembled we quantify, for the first time,
the effects of the emission of the host galaxy on the behaviour of the silicate features, a tracer of the 
properties of the hot 
dust in the torus. We rely on the classification provided by NED in order to call an object ``AGN'' 
as well as for the classification of AGN in types 1 and 2.
The sample includes a variety of AGN, from objects where 
the AGN completely dominates the MIR emission ($f_{\rm AGN} \sim 1$) to AGN 
whose MIR emission is almost entirely dominated by star formation ($f_{\rm AGN} << 1$).
We find the fraction of objects with a strong contamination from the galaxy ($f_{\rm AGN} < 0.7$) 
to be much higher among type 2 AGN (43 pre cent) than among type 1 AGN (12 pre cent). 

The emission of the host affects the behaviour of the silicate features.
Broadly speaking, the strength of the silicate feature at 9.7 \mu 
is a measure of the optical depth, $\tau_{9.7}$, along the line of sight,
and goes from emission to absorption as  $\tau_{9.7}$ increases \citep[see e.g.][Fig. 9]{fritz06}.
In this simple picture, type 1 and type 2 AGN should show the feature in emission and absorption,
respectively. Our study, however, shows that type 1 (type 2) AGN with the feature in absorption
(emission) are very common, even after subtracting the contribution of the host galaxy.
The numbers of type 1 and 2 AGN with the feature in emission increase by 20 and 50\%, respectively, 
once the host galaxy is removed, while 35\% of objects with the feature originally in absorption
exhibit it in even deeper absorption after subtraction of the host. This means that the combined spectrum 
exhibits an S$_{\rm 9.7 \,\, tot}$ intermediate between S$_{\rm 9.7 \,\, AGN}$ and that of the host.
The host galaxy nearly always shows mild silicate absorption, as the power
sources (stars) are well mixed with the absorbing dust, and therefore the strength of the silicate feature  
does not correlate with the optical depth of the gas and dust along the line of sight.
Consequently, contamination from the host increases the depth of the absorption if the feature 
intrinsic to the AGN is in emission or mild absorption, but decreases the depth (i.e. fills the gap) 
if the AGN has a deeper feature than the host.

S$_{\rm 9.7 \,\, tot}$ is scarcely ever shown in emission when the MIR emission is strongly contaminated 
by the host galaxy ($f_{\rm AGN} < 0.7$), with those objects exhibiting on 
average the feature in deeper absorption than their AGN-dominated counterparts. 
S$_{\rm 9.7 \,\, AGN}$, on the other hand, shows no dependency on $f_{\rm AGN}$ and 
appears from moderate emission to quite deep absorption with all possibilities in between. 
The lack of correlation between S$_{\rm 9.7 \,\, AGN}$ and  $f_{\rm AGN}$, two quantities
that are physically unrelated, indicates that the
decomposition mechanism successfully removes the most important part of the host galaxy
emission, allowing for an almost unbiased silicate measurement of the silicate feature in the vicinity 
of the nucleus.

We have also addressed the issue of the observed shift 
of the silicate feature at 9.7 \mu to longer wavelengths with respect to the nominal peak at 9.7 \mus.
Our decomposition method does not allow reliable estimate of $\lambda_{\rm peak}$ on the galaxy-subtracted
spectra. We therefore rely on the measurement carried out on the original IRS spectra.
We find the largest shifts ($\lambda_{\rm peak}>10.2$ \mus) to appear only in objects with an important AGN
component ($f_{\rm AGN} > 0.7$) with the feature in emission, regardless of their type. When
the feature is in absorption, and again irrespective of the type, it appears at or near 
its nominal wavelength ($\lambda_{\rm peak}<10.2$ \mus). 
Since its discovery, various scenarii have been proposed in order to explain this shift, such as the 
presence of porous dust \citep{li08, smith10}, the presence of different dust species
\citep{markwick07}, or radiative transfer effects \citep{nikutta09}. \cite{shi14}, however, posit that this is
not a radiative transfer effect but rather the effect of direct exposure of the silicates to the nuclear 
radiation, that modifies the size or the chemical composition of the grains, more along the lines of \cite{smith10}.

\subsection{Dust distribution inside and outside the torus}\label{sec:dust}

S$_{\rm 9.7 \,\, AGN}$ is seen in only moderate emission or slight absorption in most type 1 AGN, 
a behaviour that traditionally favours a clumpy morphology, both because smooth models
often predict silicates in stronger emission than observed and because, with the exception of a couple
of smooth model parameter combinations, only clumpiness can explain the feature in absorption 
in unobscured AGN. However, and even though not producing the feature in absorption,
a continuous dust distribution can also give rise to silicates in only weak
emission for a large variety of parameters, as shown in Fig. 4 of \cite{feltre12}. 
Furthermore, \cite{sirocky08} showed that the use of the \cite{ossenkopf92} dust absorption and 
scattering coefficients result in considerably less prominent silicate emission features compared 
to other dust models like \cite{draine03}. We therefore interpret the behaviour of 
S$_{\rm 9.7 \,\, AGN}$ alone as favouring the \cite{ossenkopf92} over the \cite{draine03} silicates,
but this property alone cannot provide much insight into the morphology of the dust.

The transition of the mean values of S$_{\rm 9.7}$ from weak emission to absorption from Sy1.2 to Sy1.9 can 
be explained by either dust morphologies: in a smooth medium it can be attributed to an increase 
of the inclination (as measured from the poles) and hence the intervening material, 
with the silicate feature at 9.7 \mu that arises
from the inner, hotter parts of the torus, being increasingly blocked by the
bulk of the dust as the viewing angle increases. 
In a clumpy medium, on the other hand, this could simply be attributed to different levels of obscuration
along the line of sight, independently on the orientation.

The combined strength of the silicate features at 9.7 and 18 \mu is sensitive to the chemistry and 
morphology of the dust surrounding the AGN, i.e. the torus \citep{sirocky08, thompson09,feltre12}.
To test this, we created three grids of models, two clumpy and a smooth, following
\cite{nenkova08} and \cite{feltre12}, respectively. All three sets of models share the same the primary source,
described in \cite{nenkova08}. One of the clumpy grids was created using the silicates absorption and scattering
coefficients from \cite{draine03} while the other clumpy as well as the smooth grids were created using the 
\cite{ossenkopf92} silicates. The models have been created to have {\it matched} parameters, as defined in
\cite{feltre12}, i.e. each of the smooth models in the grid has an equivalent model (in terms of geometrical 
properties) in the clumpy grid. The parameter space explored by the model grids is briefly described in
Appendix \ref{sec:models}.

Fig. \ref{fig:s10s18} shows the distributions of S$_{\rm 18 \,\, AGN}$ and S$_{\rm 9.7 \,\, AGN}$, 
compared to model predictions. The top (bottom) panel shows type 1 (type 2) AGN (in blue). 
Smooth models are shown in green, clumpy models using the \cite{ossenkopf92} and 
 \cite{draine03} silicates are shown in grey and pink, respectively. 
The spread of the observed data points indicates the variety of torus geometries in nature in terms of
size, shape and optical depths.

\begin{figure}
\begin{center}
\includegraphics[width=8.5cm,angle=0]{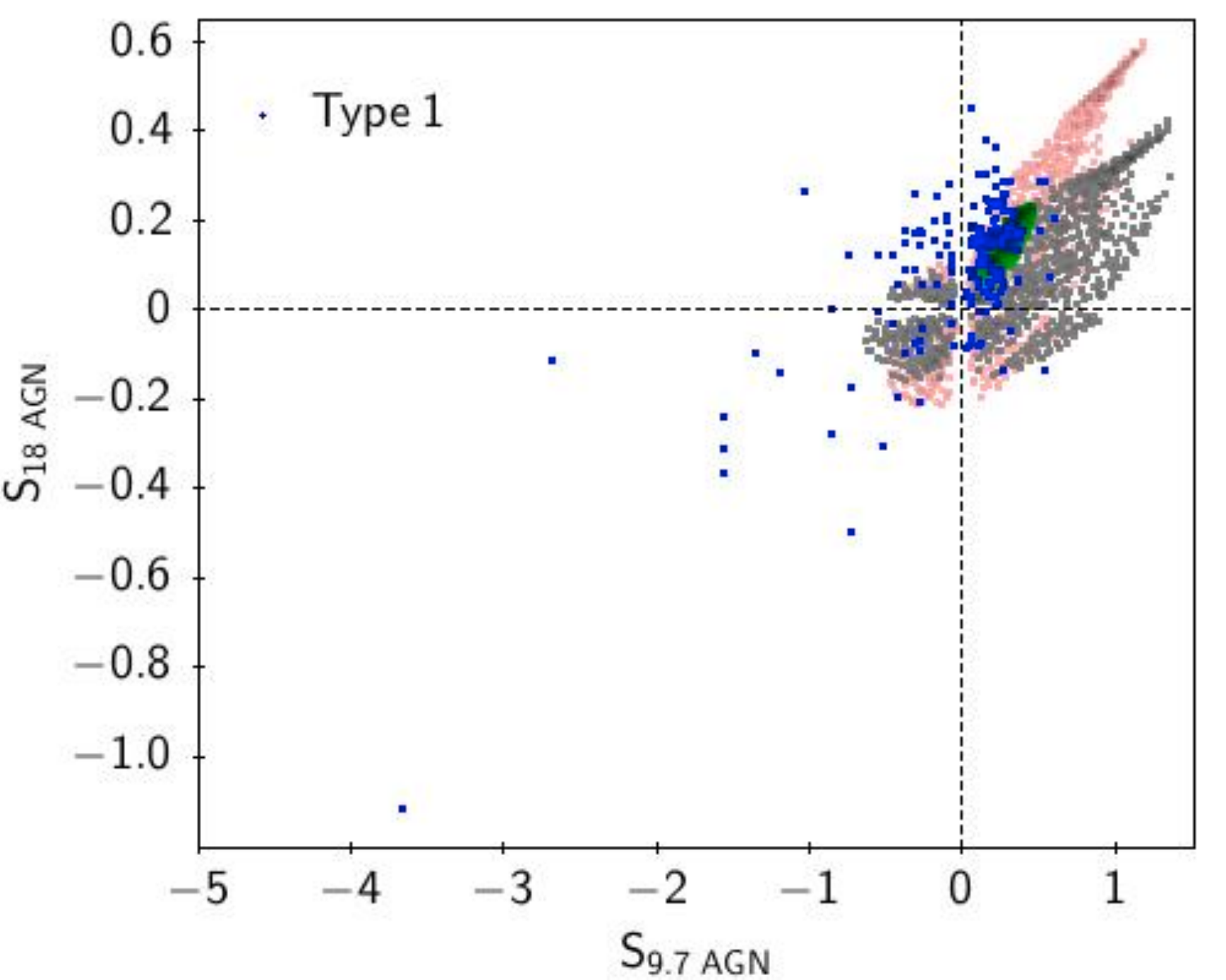}
\vskip -0.8cm
\includegraphics[width=8.5cm,angle=0]{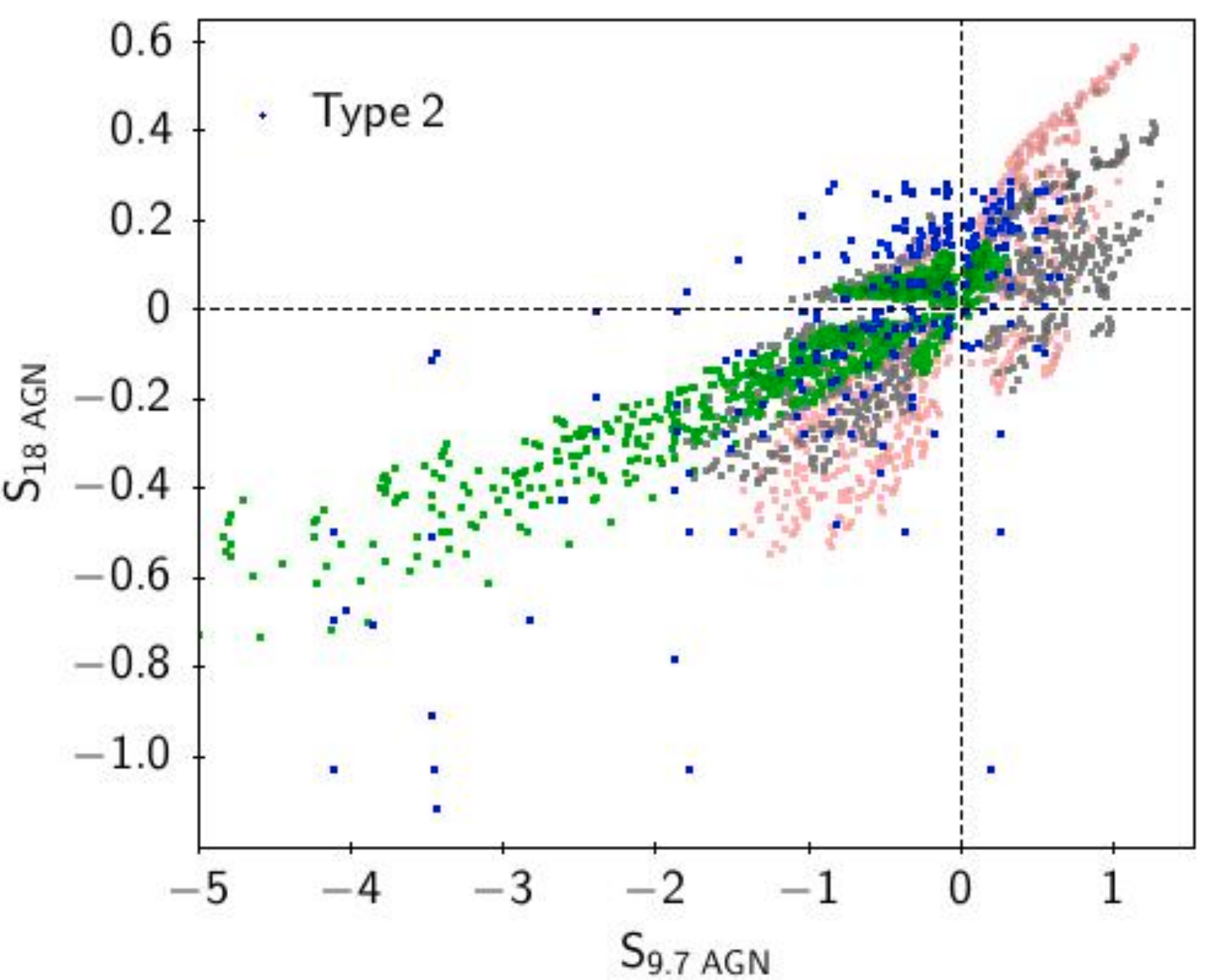}\\
\caption{S$_{\rm 18 \,\, AGN}$ as a function of S$_{\rm 9.7 \,\, AGN}$, overplotted on model predictions.
The top (bottom) panel shows type 1 (type 2) objects (in blue). 
Smooth models are shown in green, clumpy models with Ossenkopf et al. (1992)
and Draine (2003) silicates are shown in grey and pink, respectively.}
\label{fig:s10s18}
\end{center}
\end{figure}

As previously noted by other authors, the so called `astronomical' silicates \citep{draine03} produce
stronger emission features at 9.7 \mu and a wider range of S$_{\rm 18 \,\, AGN}$/S$_{\rm 9.7 \,\, AGN}$ than
what is observed, compared to the \cite{ossenkopf92} silicates. The bulk of type 1 objects lie in the
region of overlap between smooth and clumpy models. Overall, models reproduce in a better fashion the 
silicates in type 2 AGN. However, both smooth and clumpy models, albeit covering a large parameter 
space, fail to reproduce the deep absorption features seen in type 1 AGN (lower left quadrant in top panel
of Fig. \ref{fig:s10s18}). At the same time, and as already suggested by \cite{imanishi07},
a continuous dust distribution is the only morphology that can reproduce the 
deepest absorption features seen in the sample for type 2 views (lower left quadrant of bottom panel
in Fig. \ref{fig:s10s18}), at least without resorting to additional obscuration by the host galaxy.
In fact, \cite{levenson07} suggested that a deep silicate absorption feature at 9.7 \mu requires 
the primary source to be embedded in a continuous, optically and geometrically thick dusty medium, 
while a clumpy medium, with clouds illuminated from the outside, will result in a much shallower feature,
as the emission from these clouds will fill the absorption trough. 

We visually inspected all AGN with S$_{\rm 9.7 \,\, AGN} < -1.5$ for which we could find high enough
resolution images and established that they are predominantly hosted 
in galaxies with high inclinations, galaxies with very visible dust lanes crossing the centre
(e.g. NGC5793 or NGC7172), or reside in interactive systems at various stages of their interaction 
(e.g. Mkr 273, Mrk 331, NGC2623). In fact, \cite{goulding12} reach this same conclusion in their study of 20 nearby ($z<0.05$) bona-fide Compton-thick AGN. This implies that the source of the deepest silicate 
absorption features is dust in the host galaxy rather than dust in the torus \citep[see also][]{deo07}.  
In favour of this view, \cite{lagos11} find that type 1 AGN have a 
strong preference in residing in face-on galaxies, while the type 2 AGN reside in hosts of any orientation. 
The deep absorption features are, therefore, of no relevance to the modeling of the 
torus and they should not be used to favour smooth dust distributions over clumpy ones. 
We note that this is by no means an argument against the AGN 
unification scheme, but does suggest that the obscuration of the nucleus, i.e. a type 2 view, 
may, in some cases, be decoupled from the orientation of the torus.

\begin{figure}
\begin{center}
\includegraphics[width=8.5cm,angle=0]{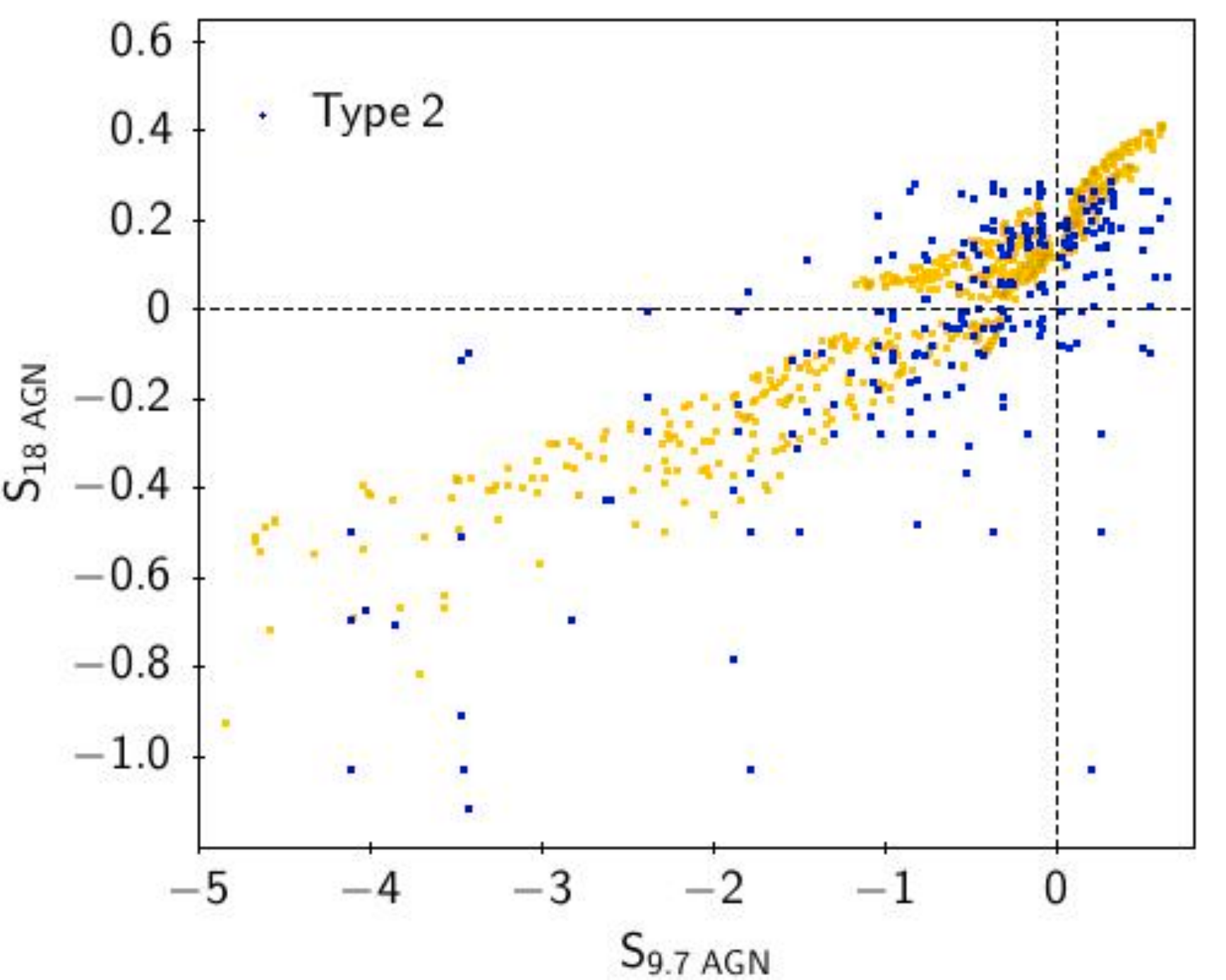}
\caption{S$_{\rm 18 \,\, AGN}$ as a function of S$_{\rm 9.7 \,\, AGN}$ for type 2 objects (in blue), 
overplotted on smooth model predictions (in yellow), using the Feltre et al. (2012) primary source.}
\label{fig:s10s18fritz}
\end{center}
\end{figure}

Finally, what is not reproduce by none of the torus models
are many of the obscured and unobscured AGN with the silicates absorption at 9.7 \mu but in 
emission at 18 \mus. Smooth models do not predict at all such a behaviour in type 1 views, while
both morphologies produce S$_{\rm 18 \,\, AGN}$ of about half the strength than that measured on the
spectra when S$_{\rm 9.7 \,\, AGN}$ is in absorption. \cite{feltre12} showed that the adopted primary source
can also affect the properties of the silicate feature. As a last resort, and since we have no control of the 
primary source of the Nenkova models, we produced an additional grid of smooth models with the
\cite{ossenkopf92} silicates but using the primary source from \cite{feltre12}, that allows for more
intrinsic AGN emission at wavelengths beyond 1 \mu compared to the primary source used by
the \cite{nenkova08} models, which we have adopted for the other grids. These models still fail to 
reproduce the objects in the upper left quadrant of the figure for type 1 views, they do however
reproduce a much larger fraction of the objects in that same quadrant for type 2 views than any of the 
other model grids presented before, as shown in Fig. \ref{fig:s10s18fritz}.

To summarise,  and even though the emerging picture is still somewhat confusing, things start clearing up:
the `cosmic' silicates \citep{ossenkopf92} represent in a satisfactory fashion the absorption 
and scattering properties of the silicates in the obscuring torus. 
Clumpiness is needed in order to produce absorption features in unobscured AGN, 
if no foreground absorber is invoked. Clumpiness can also cause the silicates to be in 
absorption at 9.7 \mu and in emission at 18 \mu in type 1 sources, but a primary source with 
more intrinsic AGN emission at $\lambda>1.0$ \mu might be necessary to create stronger 
S$_{\rm 18 \,\, AGN}$ in emission, when S$_{\rm 9.7 \,\, AGN}<0.0$.

\section*{ACKNOWLEDGMENTS}
 
The Cornell Atlas of {\it Spitzer}/IRS Sources (CASSIS) is a product of the Infrared Science 
Center at Cornell University, supported by NASA and JPL. We made use of the NASA/IPAC 
Extragalactic Database (http://ned.ipac.caltech.edu).
We used the TOPCAT software written by Mark B. Taylor (http://www.star.bris.ac.uk/~mbt/topcat/).
AHC acknowledges support by the Universidad de Cantabria Augusto Gonz{\'a}lez Linares programme 
and the Spanish Plan Nacional de Astronom{\'i}a y Astrof{\'i}sica under grant AYA2012-31447.
The research leading to these results has received funding from the European Research Council 
under the European Community's Seventh Framework Programme (FP7/2007-2013 Grant Agreement 
no. 321323).

\appendix
\section{Smooth and Clumpy model parameters}\label{sec:models}

The two original smooth and clumpy model grids cover large but not entirely overlapping parameter space.
For the purposes of this work, we {\it matched} the two sets of parameters to produce models
that can be directly comparable. Both models share the inner-to-outer radius ratio, $Y$.
The radial distribution of the dust or the clumps, $q$, can also be considered as equivalent
between the two models. For smooth models, the dust distribution in polar coordinates is given by: 
\begin{equation}
\rho\left(r,\theta \right)=\rho_0\cdot r^{-q}\cdot e^{-\gamma\times \lvert \cos(\theta)\rvert}
\end{equation}
while clumpy models assume a Gaussian angular distribution of clouds of width $\sigma$ given by:
\begin{equation}
N_{T}(\beta)=N_{0}\mbox{e}^{(-\beta^{2}/\sigma^{2})},
\end{equation}
where $N_0$ is the average number of clouds along a given radial direction on the equatorial plain
and $\beta$ (=90-$\theta$) is the angle with respect to the axis of the torus.
Considering the two extreme viewing angles for both models, the torus opening angle is of no relevance 
and $\gamma$ and $\sigma$ can be taken such to match each other. 
While smooth models use the equatorial optical depth at 9.7 \mus, $\tau_{9.7}$, as an input parameter, 
clumpy models use the optical depth of a single could calculated in the $V$-band at 0.55 \mus, $\tau_V$.
$\tau_{9.7}$ for clumpy models can be derived via the relation $\tau_{9.7} = 0.042 \times \tau_V \times N_0$,
\citep[see][]{nenkova08}.
Table \ref{tab:modparams} summarises the values of each of the parameters for the {\it matched} model grids.

\begin{table}
\begin{center}
\caption{{\it Matched} smooth and clumpy model parameters.} 
\label{tab:modparams}
\begin{tabular}{ccc}
\hline
  & Smooth & Clumpy \\
\hline
\hline
Y & 10, 30, 60, 100, 150 & 10, 30, 60, 100, 150 \\
\hline
$q$ & 0, 1 & 0, 1 \\
\hline
$\gamma$ & 1.65, 2.25, 3.24, 5.06, 9.0 & \\
$\sigma$   &  & $70^{\circ}, 60^{\circ}, 50^{\circ}, 40^{\circ}, 30^{\circ}$ \\ 
\hline
$\tau_{9.7}$ & 0.21, 0.42, 0.63, 0.84, 1.05, & \\
                       & 1.26, 1.47, 1.68, 1.89, 2.1, & \\
                       & 2.52, 2.94, 3.36, 3.78, 4.2, & \\
                       & 5.04, 5.88, 6.3, 6.72, 7.56, & \\
                       & 8.4, 10.08, 11.34, 11.76, 12.6 & \\
$N_0$          & & 1-5 \\
$\tau_V$    &  & 5, 10, 20, 30, 40, \\
                      &  & 60, 80, 100, 150 \\
\hline
\end{tabular}
\end{center}
\end{table}


\begin{thebibliography}{}
 
\bibitem[\protect\citeauthoryear{Antonucci}{1993}]{antonucci93} Antonucci R.\ 1993, ARA\&A, 31, 473
\bibitem[\protect\citeauthoryear{Buchanan et al.}{2006}]{buchanan06} Buchanan C.~L., Gallimore J.~F., O'Dea C.~P. et al.\ 2006, \aj,  132, 401
\bibitem[\protect\citeauthoryear{Clavel et al.}{2000}]{clavel00} Clavel J., Schulz B., Altieri B.et al.\ 2000, \aap, 357, 839
\bibitem[\protect\citeauthoryear{Deo et al.}{2007}]{deo07} Deo R.~P., Crenshaw D.~M., Kraemer S.~B. et al.\ 2007, \apj, 671, 124
\bibitem[\protect\citeauthoryear{Draine}{2003}]{draine03} Draine B.~T.\ 2003, \apj, 598, 1017
\bibitem[\protect\citeauthoryear{Feltre et al.}{2012}]{feltre12} Feltre A., Hatziminaoglou E., Fritz J., Franceschini A.\ 2012, \mnras, 426, 120
\bibitem[\protect\citeauthoryear{Fritz et al.}{2006}]{fritz06} Fritz J., Franceschini A., Hatziminaoglou E.\ 2006, \mnras, 366, 767
\bibitem[\protect\citeauthoryear{Goulding et al.}{2012}]{goulding12} Goulding A.~D., Alexander D.~M., Bauer F.~E. et al.\ 2012, \apj, 755, 5
\bibitem[\protect\citeauthoryear{Granato \& Danese}{1994}]{granato94} Granato G.L., Danese L.\ 1994, MNRAS, 268, 235
\bibitem[\protect\citeauthoryear{Hao et al.}{2007}]{hao07} Hao L., Weedman D.~W. Spoon H.~W.~W. et al.\ 2007, \apjl, 655, L77
\bibitem[\protect\citeauthoryear{Hao et al.}{2005}]{hao05} Hao L., Spoon H.~W.~W., Sloan G.C. et al.\ 2005, \apjl, 625, L75
\bibitem[\protect\citeauthoryear{Hatziminaoglou et al.}{2009}]{hatzimi09} Hatziminaoglou E., Fritz J., Jarrett T.\ 2009, \mnras, 399, 1206
\bibitem[\protect\citeauthoryear{Hatziminaoglou et al.}{2005}]{hatzimi05} Hatziminaoglou E., P\'erez-Fournon I., Polletta M. et al.\ 2005, \aj, 129, 1198
\bibitem[\protect\citeauthoryear{Hern\'an-Caballero et al.}{2015}]{hernan15} Hern\'an-Caballero A., Alonso-Herrero A., Hatziminaoglou E. et al\ 2015, ApJ, in press
\bibitem[\protect\citeauthoryear{Hern\'an-Caballero}{2012}]{hernan12} Hern\'an-Caballero A.\ 2012, \mnras, 427, 816
\bibitem[\protect\citeauthoryear{Hern\'an-Caballero \& Hatziminaoglou}{2011}]{hernan11} Hern\'an-Caballero A. \& Hatziminaoglou E.\ 2011, \mnras, 414, 500
\bibitem[\protect\citeauthoryear{Hern\'an-Caballero et al.}{2009}]{hernan09} Hern\'an-Caballero A., P\'erez-Fournon I., Hatziminaoglou E. et al.\ 2009, \mnras, 395, 1695
\bibitem[\protect\citeauthoryear{Hiner et al.}{2009}]{hiner09} Hiner K.~D., Canalizo G., Lacy M. et al.\ 2009, \apj, 706, 508
\bibitem[\protect\citeauthoryear{H\"onig et al.}{2006}]{hoenig06} H\"onig S.F., Beckert T., Ohnaka K. et al. \ 2006, \aap, 452, 459
\bibitem[\protect\citeauthoryear{Horst et al.}{2008}]{horst08} Horst H., Gandhi P., Smette A., Duschl W.~J.\ 2008, \aap, 479, 389
\bibitem[Houck et al.(2004)]{houck04} Houck, J.~R., Roellig, T.~L., van Cleve, J., et al.\ 2004, \apjs, 154, 18 
\bibitem[\protect\citeauthoryear{Imanishi et al.}{2007}]{imanishi07} Imanishi M., Dudley C.~C., Maiolino R. et al.\ 2007, \apjs, 171, 72
\bibitem[\protect\citeauthoryear{Lagos et al.}{2011}]{lagos11} Lagos C., Padilla N.~D., Strauss M.~A., Cora S.~A., Hao L.\ 2011, \mnras, 414, 2148
\bibitem[\protect\citeauthoryear{Lebouteiller et al.}{2011}]{lebouteiller11} Lebouteiller V., Barry D.J., Spoon H.W.W. et al.\ 2011, \apjs,  196, 8
\bibitem[\protect\citeauthoryear{Levenson et al.}{2007}]{levenson07} Levenson N.~A., Sirocky M.~M., Hao L. et al.\ 2007, \apjl, 654, L45 
\bibitem[\protect\citeauthoryear{Li, Shi \& Li}{2008}]{li08} Li M.~P., Shi Q.~J., Li A.\ 2008, \mnras, 391, L49
\bibitem[\protect\citeauthoryear{Lutz et al.}{2004}]{lutz04}  Lutz D., Maiolino R., Spoon H.~W.~W., Moorwood A.~F.~M.\ 2004, \aap, 418, 465
\bibitem[\protect\citeauthoryear{Maiolino et al.}{2007}]{maiolino07} Maiolino R., Shemmer O., Imanishi M. et al.\ 2007, \aap, 468, 979
\bibitem[\protect\citeauthoryear{Markwick-Kemper et al.}{2007}]{markwick07} Markwick-Kemper F., Gallager S.~C., Hines D.~C., Bouwman J.\ 2007, \apjl, 668, L107
\bibitem[\protect\citeauthoryear{Mason et al.}{2009}]{mason09} Mason R.~E., Levenson N.~A., Shi Y. et al.\  2009, \apjl, 693, L136
\bibitem[\protect\citeauthoryear{Mateos et al.}{2015}]{mateos15} Mateos S., Carrera F. J., Alonso-Herrero A.et al.\ 2015, \mnras, in press
\bibitem[\protect\citeauthoryear{Nenkova et al.}{2008}]{nenkova08} Nenkova M., Sirocky M.~M., Ivezi{\'c} {\v Z}., \& Elitzur M.\ 2008, \apj, 685, 147
\bibitem[\protect\citeauthoryear{Nikutta et al.}{2009}]{nikutta09} Nikutta R., Elitzur M., \& Lacy M.\ 2009, \apj, 707, 1550 
\bibitem[\protect\citeauthoryear{Noll et al.}{2009}]{noll09} Noll S.,  Burgarella D., Giovannoli E. et al.\ 2009, \aap, 507, 1793
\bibitem[\protect\citeauthoryear{Ossenkopf et al.}{1992}]{ossenkopf92} Ossenkopf V., Henning T., \& Mathis J.~S.\ 1992, \aap, 261, 567
\bibitem[\protect\citeauthoryear{Pier \& Krolik}{1992}]{pier92} Pier E.A. \& Krolik J.H.\ 1992, ApJ, 401, 99
\bibitem[\protect\citeauthoryear{Prieto et al.}{2010}]{prieto10} Prieto M.~A., Reunanen J., Tristram K.~R.~W. et al.\ 2010, \mnras, 402, 724
\bibitem[\protect\citeauthoryear{Schweitzer et al.}{2008}]{schweitzer08} Schweitzer M., Groves B., Netzer H. et al.\ 2008, \apj, 679, 101
\bibitem[\protect\citeauthoryear{Shen et al.}{2011}]{shen11} Shen Y., Richards G.~T., Strauss M.~A. et al.\ 2011, \apjs, 194, 45 
\bibitem[\protect\citeauthoryear{Shi et al.}{2014}]{shi14} Shi Y., Rieke G.~H., Ogle P.~M., Su K.~Y.~L., Balog Z.\ 2014,
\apjs, 214, 23
\bibitem[\protect\citeauthoryear{Shi et al.}{2006}]{shi06} Shi Y., Rieke G.~H., Hines D.~C. et al.\ 2006, \apj, 653, 127
\bibitem[\protect\citeauthoryear{Siebenmorgen et al.}{2005}]{siebenmorgen05} Siebenmorgen R., Haas M., Kr\"ugel E., Schulz B.\ 2005, \aap, 436, L5
\bibitem[\protect\citeauthoryear{Sirocky et al.}{2008}]{sirocky08} Sirocky M.~M., Levenson N.~A., Elitzur M., Spoon H.~W.~W., \& Armus L.\ 2008, \apj, 678, 729 
\bibitem[\protect\citeauthoryear{Smith et al.}{2010}]{smith10} Smith H.~A., Li A., Li M.~P. et al.\ 2010, \apj, 716, 490
\bibitem[\protect\citeauthoryear{Smith et al.}{2007}]{smith07} Smith J.~D.~T., Draine B.~T., Dale D.~A. et al.\ 2007, \apj, 656, 770
\bibitem[\protect\citeauthoryear{Spoon et al.}{2007}]{spoon07} Spoon H.W.W., Marshall J.A., Houck J.R. et al.\ 2007, \apj, 654, L49
\bibitem[\protect\citeauthoryear{Sturm et al.}{2006}]{sturm06} Sturm E., Hasinger G., Lehmann I. et al.\ 2006, \apj, 642, 81
\bibitem[\protect\citeauthoryear{Sturm et al.}{2005}]{sturm05} Sturm E., Schweitzer M., Lutz D. et al. \ 2005, \apjl, 629, L21
\bibitem[\protect\citeauthoryear{Thompson et al.}{2009}]{thompson09} Thompson G.~D., Levenson N.~A., Uddin S.~A., \& Sirocky M.~M.\ 2009, \apj, 697, 182 
\bibitem[\protect\citeauthoryear{Wu et al.}{2010}]{wu10} Wu Y., Helou G., Armus L. et al.\ 2010, \apj, 723, 895
\bibitem[\protect\citeauthoryear{Wu et al.}{2009}]{wu09} Wu Y., Charmandaris V., Huang, J., Spinoglio, L., Tommasin, S. 2009, ApJ, 701, 658

\end{thebibliography}
\end{document}